Astronomy
&
Astrophysics

# Cosmic-ray electron transport in the galaxy M 51

Julien Dörner[1,2], Patrick Reichherzer[1,2,3], Julia Becker Tjus[1,2], and Volker Heesen[4]

[1] Ruhr-Universität Bochum, Theoretische Physik IV: Plasma-Astroteilchenphysik, Universitätsstraße 150, 44801 Bochum, Germany
  e-mail: jdo@tp4.rub.de
[2] Ruhr Astroparticle and Plasma Physics Center (RAPP Center), Ruhr-Universität Bochum, 44780 Bochum, Germany
[3] IRFU, CEA, Université Paris-Saclay, 91191 Gif-sur-Yvette, France
[4] University of Hamburg, Hamburger Sternwarte, Gojenbergsweg 112, 21029 Hamburg, Germany

**ABSTRACT**

*Context.* Indirect observations of the cosmic-ray electron (CRE) distribution via synchrotron emission is crucial for deepening the understanding of the CRE transport in the interstellar medium, and in investigating the role of galactic outflows.
*Aims.* In this paper, we quantify the contribution of diffusion- and advection-dominated transport of CREs in the galaxy M51 considering relevant energy loss processes.
*Methods.* We used recent measurement from M 51 that allow for the derivation of the diffusion coefficient, the star formation rate, and the magnetic field strength. With this input, we solved the 3D transport equation numerically including the spatial dependence as provided by the measurements, using the open-source transport framework CRPropa (v3.1). We included 3D transport (diffusion and advection), and the relevant loss processes.
*Results.* We find that the data can be described well with the parameters from recent measurements. For the best fit, it is required that the wind velocity, following from the observed star formation rate, must be decreased by a factor of 5. We find a model in which the inner galaxy is dominated by advective escape and the outer galaxy is composed by both diffusion and advection.
*Conclusions.* Three-dimensional modelling of cosmic-ray transport in the face-on galaxy M51 allows for conclusions about the strength of the outflow of such galaxies by quantifying the need for a wind in the description of the cosmic-ray signatures. This opens up the possibility of investigating galactic winds in face-on galaxies in general.

## 1. Introduction

In this age of multi-wavelength and multi-messenger astronomy, nearby galaxies that allow for spatially resolved substructures can be observed at different wavelengths in such detail that there is differential information about the population of basically all of the ingredients needed to describe spatially resolved cosmic-ray transport and interaction of cosmic rays. In particular, information on the 3D magnetic field structures, a differential view on the star formation rate and secondary properties such as the spectral index of cosmic rays, the cosmic-ray diffusion coefficient, as well as the advection velocity of the plasma can be provided for both edge-on (Heesen et al. 2018; Miskolczi et al. 2019; Heald et al. 2022) and face-on (Murphy et al. 2008; Tabatabaei et al. 2013; Mulcahy et al. 2016) galaxies. This wealth of data brings 1D transport models beyond their limits.

One-dimensional models try to either describe a galaxy in edge-on or face-on geometry. Clearly, while this is a useful simplification, it neglects the true 3D structure of a galaxy. For the edge-on case, one hence neglects the local concentration of star formation, such as in spiral arms where advection is more important than diffusion. Also, the conservation of magnetic flux means that at least extensions to quasi 1D models are required in order to lead to a decreasing magnetic field strength, such as the 'flux tube' model (Heald et al. 2022). On the other hand, in face-on galaxies the radio continuum emission is smeared out in comparison with the star formation, so that this map can be convolved to either minimize the difference between the two maps (Murphy et al. 2008) or linearize the cross-correlation between them (Berkhuijsen et al. 2013). This subsequently provides an estimate of the cosmic-ray diffusion length. A shortcoming of this method is that it neglects the influence of the radio halo along the line of sight. Also, as Mulcahy et al. (2016) have shown, cosmic-ray electrons (CREs) escape from the galaxy and this cannot be neglected, but it is indeed required in order to explain the radio continuum spectrum. The model by Mulcahy et al. (2016) was a first attempt to move beyond purely descriptive work in face-on galaxies, solving the diffusion-loss equation to model the radio spectral index. Obviously, the most promising way would be to go beyond this simplification and describe a galaxy with both cosmic-ray diffusion and advection in a 3D model. In this paper, we use the publicly available Monte-Carlo code CRPropa (Batista et al. 2016; Merten et al. 2017; Alves Batista et al. 2022) to describe the 3D transport in M 51. While originally written to describe the extragalactic transport of hadronic cosmic rays via the solution of the equation of motion, CRPropa has been extended to a second propagation method, that is solving the transport equation via the approach of stochastic differential equations (SDEs). The conversion of a Fokker-Planck equation into an SDE is useful here, as the particle densities are derived from the pseudo-particle trajectories. This way, the equation of motion approach and the transport equation can be treated in one framework since both work with individual particle trajectories (Merten et al. 2017). This approach also allows for both continuous and catastrophic losses for the production of full particle showers in the interactions that can be followed up on, etc.

Modelling the 3D transport of CREs in M 51 mainly depends on the following three assumptions: (1) the diffusion coefficient and its energy scaling, (2) the escape scale height for CREs, and (3) the advection speed profile. We subsequently implement these properties in the transport code as described in the following Sect. 2 and fit the parameters to the observed properties. The results are discussed in Sect. 3 and conclusions are made in Sect. 4.

## 2. Transport model

The transport of CREs can be described by the diffusion-advection equation (e.g. Becker Tjus & Merten 2020)

$$\frac{\partial n}{\partial t} = D\,\nabla^2 n - \boldsymbol{v}\cdot\nabla n - \frac{\partial}{\partial E}\left[\frac{\partial E}{\partial t}\,n\right] + S, \quad (1)$$

assuming isotropic diffusion, where $n$ is the particle density distribution, $D$ is the diffusion coefficient, $\partial E/\partial t$ is the energy loss term described in Sect. 2.1, $S$ is the source term, and $\boldsymbol{v}$ is the advection speed derived from the Star Formation Rate surface Density (SFRD) as described in Sect. 2.2. Here, no leakage term is included, as in our 3D simulations all particles that reach the boundary of the galaxy (see Sect. 2.5) leave the simulation volume.

### 2.1. Energy loss

Energy loss terms $\partial E/\partial t$ for synchrotron emission and inverse Compton scattering are taken into account in the CRE transport equation. In our simulation, we applied the energy loss as a continuous process following the parametrization given by Mulcahy et al. (2016)

$$\frac{\partial E}{\partial t} = 8\times 10^{17}\,E^2\left(U_{\rm rad} + 6\times 10^{11}\,\frac{B^2(\boldsymbol{r})}{8\pi}\right)\frac{\rm GeV}{\rm s}, \quad (2)$$

where $E$ is the CRE energy in GeV, $U_{\rm rad} = 1$ eV is the energy density of the interstellar radiation field, and $B(\boldsymbol{r})$ is the root mean square of the magnetic field. The magnetic field model was implemented to decrease exponentially with the height $z$ above the scale height $h_b$:

$$B(\boldsymbol{r}) = B(r_{\rm gc})\cdot\exp\left\{-\frac{|z|}{h_b}\right\}. \quad (3)$$

Here, $r_{\rm gc}$ denotes the galactocentric radius, and $z$ is the height above the galactic plane. The radial profile of the magnetic field strength was measured by Heesen et al. (2022) and shown in Fig. 1. The exponential cutoff scale $h_b$ is listed in Table 1.

### 2.2. Source distribution

The acceleration of CREs is believed to happen in star forming regions, possibly at the shock front of supernova remnants; interested readers can refer to Becker Tjus & Merten (2020), for example, for a review. Diffusive reacceleration in these regions is possible as well (Tolksdorf et al. 2019). Other possible sources are pulsar wind nebula (López-Coto et al. 2022). While the different acceleration scenarios can influence the spectral energy distribution of the sources, the origin for all of them are the star

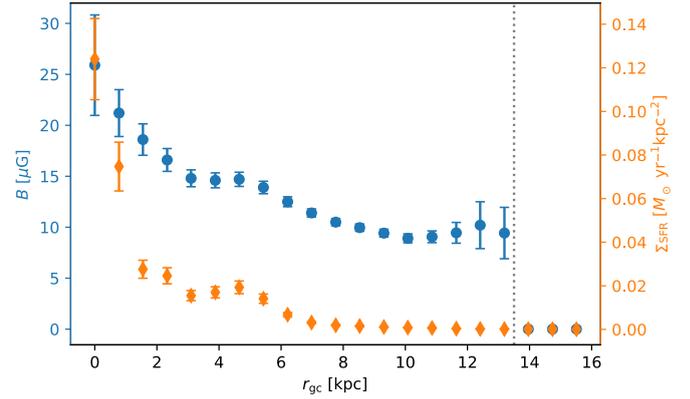

**Fig. 1.** Radial dependence of the root mean square of the magnetic field strength (blue) and the Star Formation Rate surface Density (SFRD, orange). Data are taken from Heesen et al. (2019, 2022). The grey dotted line indicates the restriction of our model.

**Table 1.** Parameters for the magnetic field scale height $h_b$ and the height of the disc $h_d$.

| Model | $h_b$ | $h_d$ |
|---|---|---|
| A | 7 kpc | 7 kpc |
| B | 6 kpc | 3 kpc |
| C [a] | $h = h(r_{\rm gc})$ | |

**Notes.** [a] Variable scale heights as a function of the galactocentric radius as presented in Mulcahy et al. (2016).

formation regions. Therefore, we assume that the radial source position of the electrons follows the observed SFRD (see Fig. 1). The SFRD was estimated using a hybrid star formation map from a combination of GALEX 156-nm far-ultraviolet and *Spitzer* 24-μm mid-infrared data (Leroy et al. 2008).

### 2.3. Cosmic-ray advection

Even the strength of the galactic wind is assumed to be proportional to the SFRD. This is motivated both by a similarity analysis of planar blast waves (Vijayan et al. 2020) and radio continuum observations of radio haloes in edge-on galaxies (Heesen 2021). The galactic wind speed as measured from ionized gas depends on the SFRD. Hence, we used this as our parametrization.

We took a galactic wind in $z$-direction $\boldsymbol{v}(\boldsymbol{r}) = \text{sgn}(z)\,v(r_{\rm gc})\,\boldsymbol{e}_z$, where

$$v(r_{\rm gc}) = 10^{3.23}\left(\Sigma_{\rm SFR}(r_{\rm gc})\right)^{0.41}\,{\rm km\,s^{-1}} \quad (4)$$

is the best-fit wind velocity following the radial-dependent SFRD found in Heesen et al. (2018) and sgn denotes the sign function. In this model, the wind velocity does not depend on the $z$ position. In galactic wind models, the wind speed is zero in the galactic disc and then with height. Depending on the assumptions of geometry and energy and mass injection, this acceleration can be either gradual over a few kiloparsecs (Everett et al. 2008) or rapid over a few hundred parsecs (Yu et al. 2020). So far, no consensus has been reached as to the vertical acceleration profile either, as the properties of the CRE distribution and magnetic field strength are difficult to disentangle (Heesen 2021). Hence, we make the simplifying assumption of a constant wind speed.

**Table 2.** Parameters and modules for the simulation in CRPropa3.1.

|  | Module | Parameters |
|---|---|---|
| Propagation | `DiffusionSDE` | $l_{min} = 0.1$ pc, $l_{max} = 10$ kpc |
| Observer | `ObserverTimeEvolution` | $N_{step} = 1000$, $\Delta T = 500$ kyr |
| Boundary | `MinimumEnergy` | $E_{min} = 0.1$ GeV |
|  | `MaximumTime` | $T_{max} = 2.5$ Gyr |

**Notes.** Here, $l_{min}$ and $l_{max}$ denote the range of the propagation steps and $T_{max}$ is the maximal simulation time until a stationary solution was reached.

### 2.4. CRE diffusion

Deflections of CREs in the Galactic magnetic field introduce diffusive transport behaviour, which is characterized by the diffusion tensor $\hat{D}$ that enters the transport equation. It is known that in galaxies, spatial diffusion can be anisotropic- or isotropic-dependent on the environment (Sampson et al. 2022). In the absence of detailed knowledge of the 3D magnetic field structure for M 51, we assume scalar isotropic diffusion. We note that a more realistic, 3D modelling of the diffusion tensor requires knowledge of the relation between parallel and perpendicular diffusion coefficient components (Reichherzer et al. 2022). Assuming that the magnetic field lines are mainly orientated in the galactic plane, the escape would be dominated by perpendicular diffusion. In this case, by choosing an isotropic diffusion, the transport in the plane would be underestimated. However, parallel escape is suppressed by the geometry of the large, flat galactic disc anyways (see Sect. 2.5) and we probably suppress this component a bit more by isotropic diffusion.

The diffusion coefficient dependency on the parameters of the CREs and the environment relies on the dominant scattering mechanism. Recent observational data, for example, spectra of positrons and their parent protons in the Milky Way (Cowsik & Huth 2022) or analytical considerations employing advection-dominated escape models (Recchia et al. 2016), suggest energy-independent diffusion coefficients of charged particles up to several GeV. This relates to gyroradii of the order of $10^{-7}$ pc in approximately µG magnetic fields shown in Fig. 1[1]. However, even for diffusion-dominated escape, various explanations exist for energy-independent diffusion (see e.g. Kempski & Quataert 2022; Cowsik & Huth 2022 and references therein). Possible explanations include resonant scattering of CREs by self-excited fluctuations when these waves are excited through the streaming instability in the absence of damping (Kempski & Quataert 2022). Energy-independence can also be achieved for particle scattering in pre-existing magnetohydrodynamic turbulence (Cowsik & Huth 2022) or through the influence of the Parker instability, causing the leakage of cosmic rays out of the galaxy (Parker 1966, 1969). Also, the field-line-random walk that may contribute to perpendicular diffusion at these energies exhibits energy-independent diffusion (Minnie et al. 2009; Reichherzer et al. 2020). Regardless of which of the effects or combination described above holds, we assume energy-independent diffusion. In lack of a theoretically motivated diffusion coefficient, we compared the observation-based diffusion coefficient $D \approx$ $2 \times 10^{28}$ cm$^2$ s$^{-1}$ (Heesen et al. 2019) with the best-fit model from Mulcahy et al. (2016). We compared our result to energy-dependent diffusion, which led to a significantly worse fit for the data (see Appendix B). Such a behaviour is also suggested by the 1D diffusion models of Mulcahy et al. (2016) and the convolution experiments of Heesen et al. (2019).

### 2.5. Geometry of M51

To model the geometry of M51, we took a cylindrical form of the galaxy with a maximal radius of $R_{max} = 15$ kpc and a height $h_d$, allowing a $z$-position $-h_d \leq z \leq h_d$. The parameter for $h_d$ is not fully known. Therefore, we present the results for three different models as follows: Model A considers a large-scale height for the galactic height and for the magnetic field of $h_d = h_b = 7$ kpc. This value is not realistic, but it was chosen to see the impact of the parameter. Model B is based on the observed synchrotron emission scale height of 1.5 kpc (Krause et al. 2018). Therefore, the height of the disc is considered to be $h_d = 3$ kpc and, for the magnetic field height, we used $h_b = 6$ kpc. Model C follows the variable scale height presented in Mulcahy et al. (2016). Here, a scale height reads as

$$h(r_{gc}) = \begin{cases} 3.2 \text{ kpc}, & r_{gc} \leq 6 \text{ kpc} \\ 3.2 \text{ kpc} + \frac{5.6}{6}(r_{gc} - 6 \text{ kpc}), & \text{else} \\ 8.8 \text{ kpc}, & r_{gc} \geq 12 \text{ kpc} \end{cases}. \quad (5)$$

All model parameters are summarized in Table 1.

### 2.6. Simulation setup

To solve the transport equation, we used the method of stochastic differential equations (SDEs) implemented in CRPropa 3.1 (Batista et al. 2016; Merten et al. 2017). We simulated $10^5$ pseudo-particles in the energy range of 0.1 GeV to 50 GeV. We assumed a source with an injection $dN/dE|_{source} \propto E^{-2}$. The CRE density distribution was considered for 1000 time steps up to 500 Myr to calculate the stationary solution of the transport equation (Eq. (1)) following Merten et al. (2017). All particles reaching the boundary have been lost to the intergalactic medium. Also, particles propagating longer than the maximum simulation time $T_{max}$ were taken out of the simulation. Here, $T_{max} = 2.5$ Gyr is just an assumption. In Mulcahy et al. (2016), it is shown that the CRE distribution reaches a steady state after 500 Myr. Therefore, our choice of a 5 times higher simulation time is more conservative. The details of the used modules and given parameters for the simulation are given in Table 2. We analysed the CRE spectrum in a slightly smaller energy range than simulated to minimize numerical artefacts. The range of the power law fits is $0.5 \leq E/\text{GeV} \leq 6$.

---

[1] We note that the gyroradii are many orders smaller than the correlation length $\gtrsim 1$ pc of the turbulence. This transport regime is different compared to the regime often discussed in heliospheric studies, where cosmic-ray protons exhibit energy-dependent diffusion $\gtrsim 10$ MeV (Forman & Gleeson 1975; Matthaeus et al. 2003; Fraschetti & Giacalone 2012). In this case, the gyroradius is not much smaller than the correlation length, by a few orders at most.

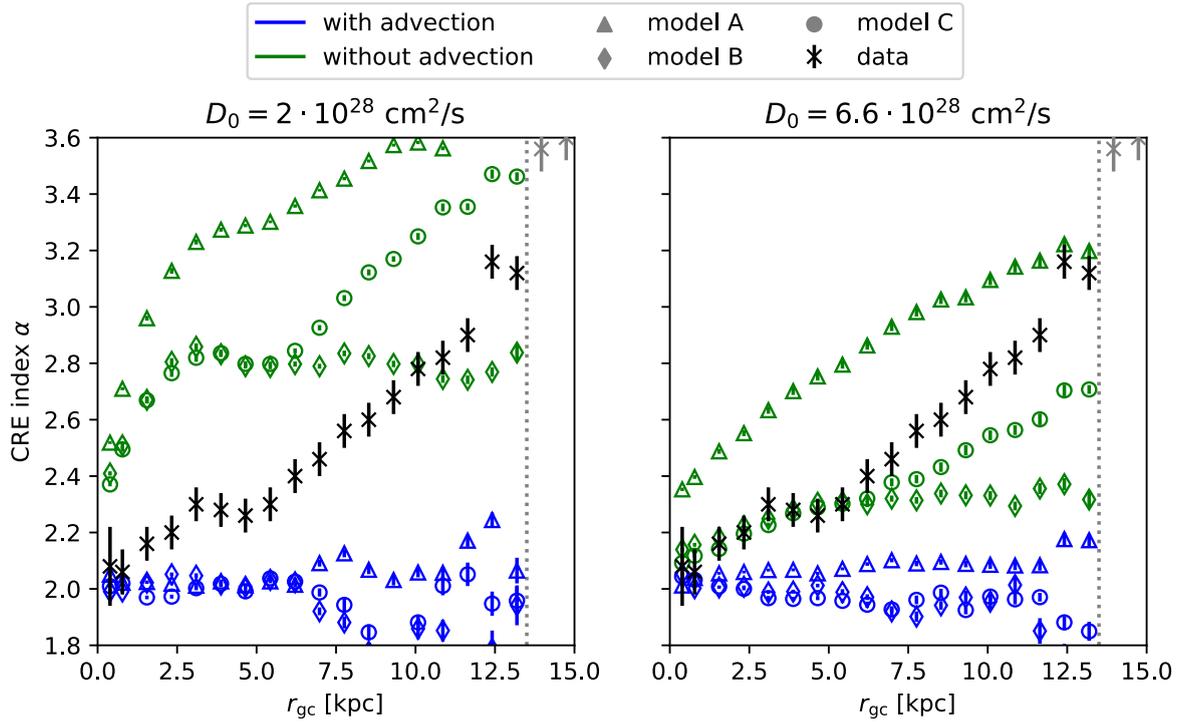

**Fig. 2.** CRE spectral index as a function of the galactocentric radius. The left panel shows the simulation results using the measured diffusion coefficient from Heesen et al. (2019) and the right panel uses the best-fit value from Mulcahy et al. (2016). The model parameters are shown in Table 1. Green points indicate simulations without advection and blue point show those with advection. The data are taken from Heesen et al. (2019).

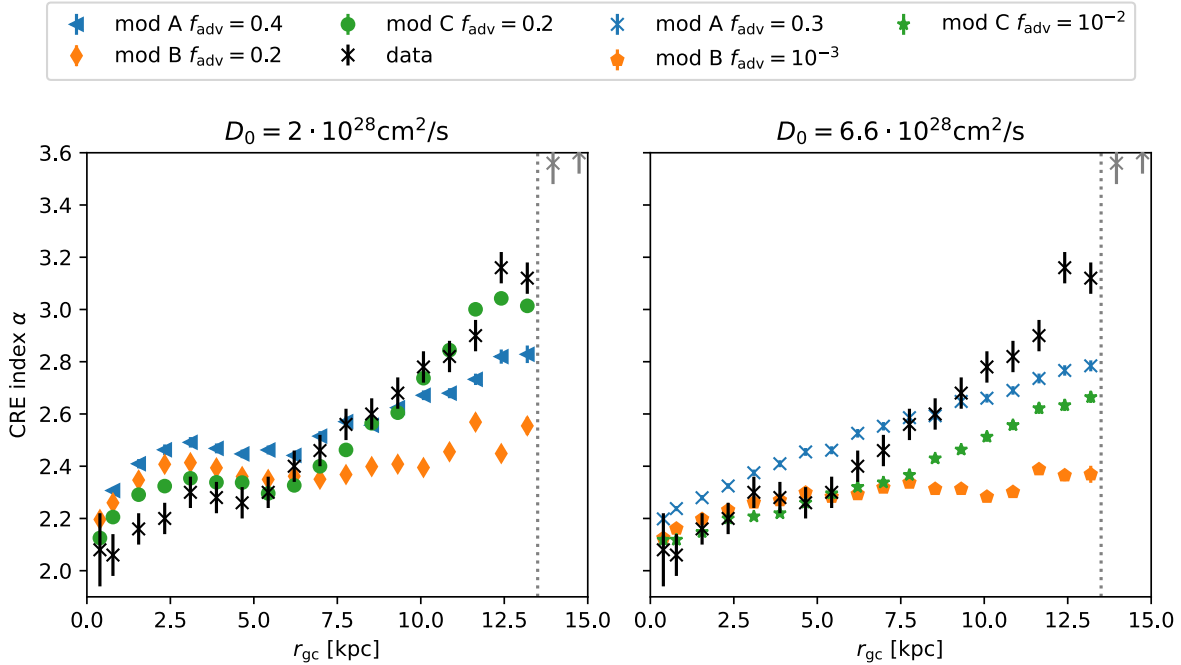

**Fig. 3.** Radial variation of the CRE spectral index using the optimized wind speed. The model parameters are shown in Table 1. The left panel shows the observed diffusion coefficient from Heesen et al. (2019) and the right panel shows the best-fit value for the diffusion coefficient from Mulcahy et al. (2016).

## 3. Results

Taking the model as described before, the resulting CRE spectral index is presented in Fig. 2, where the model without advection (green points) and including advection as described in Sect. 2.2 (blue points) is compared for two diffusion coefficients. In the case of the lower diffusion and neglecting advection, the spectra for all models are too steep due to the high retention time and corresponding high energy loss. Only model B undershoots the observed data in a range slightly, but it does not show the correct radial behaviour. In the case of higher diffusion, model C (green circle) fits the data in the inner galaxy ($r_{gc} < 6$ kpc) well. Only in the outer part of the galaxy can a difference between the data and the model be seen. This is due to different data for the magnetic field strength, star formation rate, and spectral synchrotron index in Mulcahy et al. (2016). Another reason for the difference could be the difference between the 1D diffusion model used by Mulcahy et al. (2016) and the 3D approach in this work.

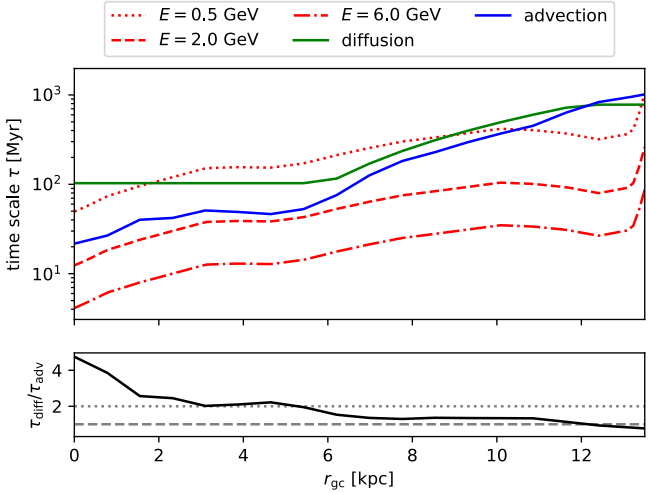

**Fig. 4.** Determining the dominant timescale. *Upper panel*: timescales for the escape via diffusion (green line) and advection (blue line) to the z-direction. Additionally, the energy loss timescale is given for three different energies (red lines; the line style denotes energy). *Lower panel*: ratio between the advection and diffusion timescale.

In the case where advection is taken into account (Fig. 2, blue points), the observed CRE spectral index is near the injection spectrum $\propto E^{-2}$. This is due to high advection speed and a quick loss of all particles. The case of the high variation for the low diffusion models in the outer galaxy can be explained by the low number of observed pseudo-particles in this domain. In this part of the galaxy, the SFRD is so low that there is nearly no production of high energy CREs. But due to the high advection speed, the particles leave the simulation volume before they can diffuse in the outer galaxy.

Following these observation, a galactic wind significantly weaker than indicated by the SFRD is necessary to match the observed data. Taking this into account, we introduced a scaling factor $f_{adv}$ in Eq. (4) and optimized this value to fit the data best. Details on the optimization are given in Appendix A. The final CRE indices using the optimal value for the advection normalization are shown in Fig. 3. It can be seen that the lower diffusion coefficient shows a significantly better fit to the data. The best fit provides model C. In this case, the optimal normalization factor is $f_{adv} = 0.2$. The models with a constant scale height (model A and model B) do not fit the radial gradient. In the inner galaxy ($r_{gc} \lesssim 7$ kpc) the CRE spectra are too steep and in the outer galaxy too flat.

Taking the geometry of model C and the lower diffusion coefficient as the best-fit model, analysing the timescales shows the dominant processes. Here, the diffusion timescale is defined as $\tau_{diff} = h(r_{gc})^2/D_0$ and the advection timescale as $\tau_{adv} = h(r_{gc})/v(r_{gc})$. In Fig. 4 it is shown that the escape inner galaxy ($r_{gc} \lesssim 7$ kpc) is dominated by advection. The escape in the outer galaxy is composed by both diffusion and advection. In the relevant energy range ($E > 2$ GeV), the energy loss time is much shorter than the escape timescale. This leads to a steepening of the CRE spectrum. The rise of the energy loss timescales (red lines) at the edge of the galaxy is due to the vanishing magnetic field strength in the outer part (see Fig. 1). This region is excluded in our analysis.

## 4. Conclusions

Our best-fit model to the radial gradient of the observed CRE spectra has the following settings:
1. The diffusion coefficient is independent of the energy with $D_0 = 2 \times 10^{28}$ cm$^2$ s$^{-1}$. This result is in agreement with the measurement from Heesen et al. (2019).
2. The scale height for the escape of CREs depends on the galactocentric radius. We used $h_d = 3.2$ kpc for the inner galaxy ($r_{gc} \leq 6$ kpc) and increased it linearly up to $h_d = 8.8$ kpc at $r_{gc} = 12$ kpc.
3. The advection speed following the SFRD was reduced by a factor of 5 compared to the measurements in Heesen et al. (2018). The discrepancy can possibly be explained by the fact the radio continuum observations use global SFRD values with $\Sigma_{SFR} = SFR/(\pi r_\star^2)$, where $r_\star$ is the radial extent of the star-forming disc. If the wind is launched from the central area of the galaxy, the SFRD would be correspondingly higher if one were to use an effective radius $r_e \approx r_\star/2$; this would reduce the advection speed normalization in Eq. (4) by a factor of 2. While these advection speeds may still be slightly too high, the wind velocities of the ionized gas as measured by Heckman & Borthakur (2016) are in fair agreement with our new results.

We conclude that the escape of CREs is governed by different mechanisms in the inner and outer part of M 51: the inner galaxy ($r_{gc} \leq 7$ kpc) appears as an advection-dominated region; whereas, in the outer galaxy, both diffusion and advection have to be taken into account. This is basically consistent with the picture of a wind being present. In contrast to previous results, however, our best-fit model results in a wind that is a factor of 5 smaller than derived indirectly from the star-formation rate. Finally, we can show here that with a 3D transport model, it is possible to constrain the propagation environment of CREs, concerning diffusion and advection. More specifically, the 3D modelling represents an additional way of indirectly deducing the strength of a wind velocity of the face-on galaxy M 51, opening the possibility to systematically investigate galactic winds for face-on galaxies in general.

*Acknowledgements.* J.D., P.R., and J.B.T. acknowledge the support from the Deutsche Forschungsgemeinschaft, DFG via the Collaborative Research Center SFB1491 Cosmic Interacting Matters – From Source to Signal. Part of this work was supported by the DFG project number Ts 17/2–1. This work was made possible by the following software packages: CRPropa (Batista et al. 2016; Merten et al. 2017; Alves Batista et al. 2022), dask (Rocklin 2015), ipython (Pérez & Granger 2007), matplotlib (Hunter 2007), numpy (Harris et al. 2020) and pandas (Wes McKinney 2010).

## Appendix A: Optimization of the wind speed

Starting with the simulation results from the best-fit wind velocity in fig. 2, it can be seen that the escape due to advection is far too high to match the observed data. Therefore, we introduced a normalization factor $f_{adv}$ in eq. 4, so the advection velocity reads as

$$v(r) = f_{adv} \cdot 10^{3.23} \text{ km/s} \, (\Sigma_{SFR})^{0.41} \, e_z \quad . \quad (A.1)$$

To quantify the quality of the fit of the simulation results to the data, a $\chi^2$ variable

$$\chi^2_{red} = \frac{1}{df} \sum_{i=1}^{N} \frac{(\alpha_{sim,i} - \alpha_{obs,i})^2}{\sigma_{obs,i}} \quad (A.2)$$

was used, where $\alpha_i$ denotes the spectral index in the $i$-th bin, $\sigma_i$ is the error of the observed data, and $df = N - 1$ is the degree of freedom. Here only data for $r_{gc} < 13.5$ kpc are taken into account, because the magnetic field vanishes for higher values (see fig. 1).

Trying to fit the observed data best, we tried values for $f_{adv}$ between 0.1 and 1 in steps of $\Delta f = 0.1$. Additionally, we tested $f_{adv} = 10^{-2}, 10^{-3}, 10^{-4}$ to compare the expected behaviour for low normalization factors.

The results are shown in fig. A.1 for all three models and both diffusion coefficients. It can be seen that **model C** leads to a sharp global minimum. To reach a minimized $\chi^2_{red} \leq 1$, only a small deviation in the wind speed fraction of about $f_{adv} = 0.2 \pm 0.1$ is allowed. This can also be seen in fig. A.2 where the spectral index is plotted against the galactocentric radius for different advection speed fractions $f_{adv}$. Only the values of $f_{adv} = 0.2 \pm 0.1$ are in reasonable agreement with the data (black cross).

The simulation for the higher diffusion coefficient $D_0 = 6.6 \cdot 10^{28}$ cm$^2$/s does not show a plausible minimum for **model A** or **model B**. This is expected due to the fact that the value of the diffusion coefficient is fitted to match the escape timescale for the CREs in Mulcahy et al. (2016). Therefore, any additional contribution of a galactic wind would lead to shorter escape timescales and a flatter spectrum. The best-fit values for the wind speed fraction are shown in table A.1.

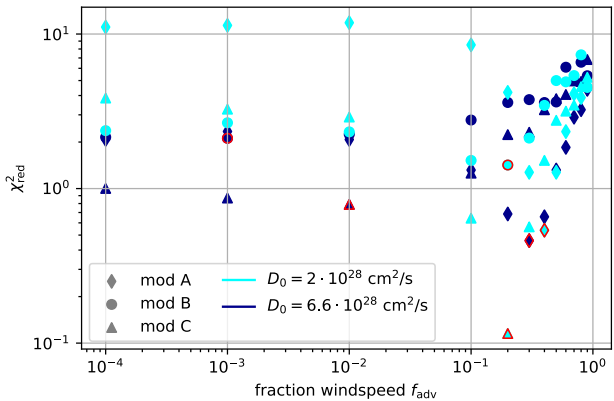

Fig. A.1: Optimization value $\chi^2_{red}$ depending on the fraction of the wind speed $f_{adv}$. The minimal value is marked with a red edge of the data point.

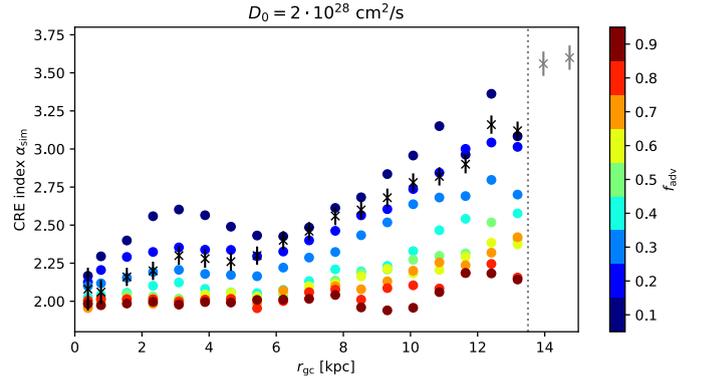

Fig. A.2: CRE spectral index for different wind speed $f_{adv}$ (colour-coded) in the **model C**.

| Diffusion coefficient | model A | model B | model C |
|---|---|---|---|
| $2 \cdot 10^{28}$ cm$^2$/s | 0.2 | 0.4 | 0.2 |
| $6.6 \cdot 10^{28}$ cm$^2$/s | $10^{-2}$ | $10^{-3}$ | 0.3 |

Table A.1: Optimal value for the normalization of the wind speed $f_{adv}$ in the different models.

## Appendix B: Energy-dependent diffusion

Although the observation indicates an energy-independent diffusion coefficient, we compared our model taking an energy dependence into account. In this appendix, we restrict our geometry to only the best-fit model **[model C]**, with the radial-dependent scale height. The optimization of the wind speed was performed as presented in appendix A. We compared different diffusion models to the observed data. The first model assumes a diffusion coefficient similar to the observation in the Milky Way, assuming a Kolmogorov-like turbulence scaling the diffusion reads as

$$D^I(E) = 6.1 \cdot 10^{28} \text{ cm}^2/\text{s} \, (E_{4\text{GeV}})^{\frac{1}{3}} \quad , \quad (B.1)$$

where $E_{4\text{GeV}}$ is the energy in units of 4 GeV. As a second comparison, we normalized the diffusion coefficient to the observed value of $D_0 = 2 \cdot 10^{28}$ cm$^2$/s taking the same energy scaling as before. In this case, the diffusion coefficient reads as

$$D^{II}(E) = 2 \cdot 10^{28} \text{cm}^2/\text{s} \, (E_{4\text{GeV}})^{\frac{1}{3}} \quad . \quad (B.2)$$

The resulting CRE spectra are shown in fig. B.1. Both cases fit the data significantly worse. Especially in the inner galaxy, the flat spectra cannot be reproduced by the energy-dependent diffusion.

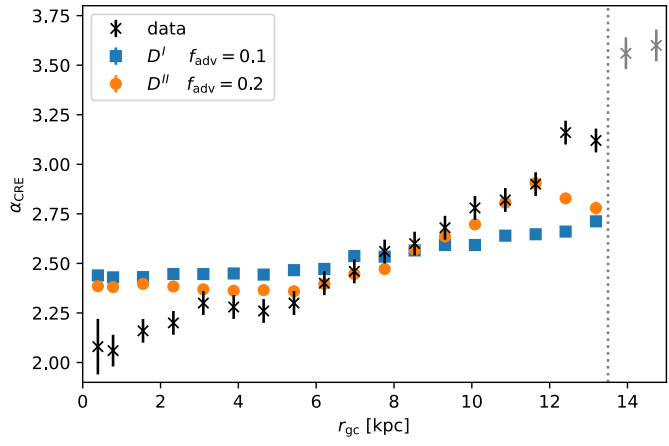

Fig. B.1: CRE spectral index for the energy-dependent diffusion models using the geometry of **model C**.